\documentclass{llncs}
\usepackage{cite}
\usepackage{graphicx}

\let\oldendproof\endproof
\def\endproof{\qed\oldendproof}

\begin{document}
\mainmatter
\title{Trees with Convex Faces and Optimal Angles}
\author{Josiah Carlson \and David Eppstein}
\institute{Computer Science Department, University of California, Irvine\\
\email{\{jcarlson,eppstein\}@uci.edu}
}

\maketitle

\begin{abstract}
We consider drawings of trees in which all edges incident to leaves can be extended to infinite rays without crossing, partitioning the plane into infinite convex polygons.  Among all such drawings we seek the one maximizing the angular resolution of the drawing.  We find linear time algorithms for solving this problem, both for plane trees and for trees without a fixed embedding.  In any such drawing, the edge lengths may be set independently of the angles, without crossing; we describe multiple strategies for setting these lengths.
\end{abstract}

\section{Introduction}

Suppose we wish to draw a tree in the plane by first setting the slopes of its edges, and then independently setting the edge lengths. For what choices of slopes are we guaranteed that the drawing will be non-self-crossing, no matter what lengths are chosen?

\begin{figure}[p]
\centering\includegraphics[width=1.9in]{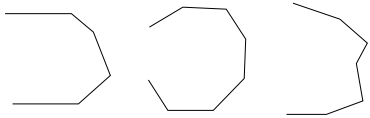}
\caption{Left: a convex arch. Center: not a convex arch, as the angles span a range larger than $\pi$. Right: not a convex arch, as the angles of the segments are not in sorted order.}
\label{fig:arches}
\end{figure}

\begin{figure}[p]
\centering\includegraphics[width=1.4in]{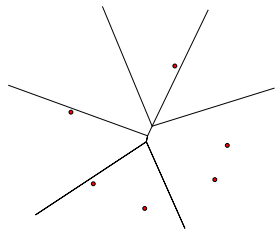}
\caption{The farthest point Voronoi diagram partitions the plane into infinite polygonal cells, much like the partition formed by extending the leaf edges of a tree drawing with convex faces.}
\label{fig:farvd}
\end{figure}

\begin{figure}[p]
\centering\includegraphics[width=3.75in]{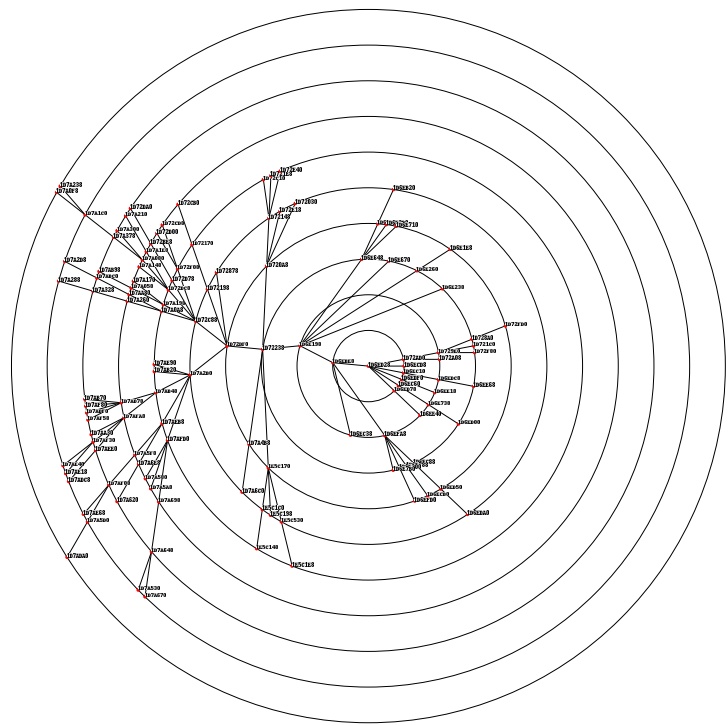}
\caption{A tree drawn with convex faces}
\label{fig:example}
\end{figure}

To answer this question, we define a \emph{convex arch} to be a polygonal chain spanning a range of angles of at most $\pi$, so that the edges occur on the chain in sorted order by their angles within that range (Figure~\ref{fig:arches}). We say that a tree drawing has \emph{convex faces} if the path between all consecutive pairs of leaves (in the radial order of leaves around the root of the tree) is a convex arch.

If a tree drawing has convex faces, then the edges incident to the leaves may be extended to infinite rays, transforming each arch into an infinite convex polygon. These polygons partition the plane, similarly to the partition formed by a farthest neighbor Voronoi diagram (Figure~\ref{fig:farvd}; see also \cite{LioMei-CGTA-03} for Voronoi related tree drawing algorithms). There can be no crossings within any of these convex faces, so the drawing is planar, regardless of the drawing's edge lengths. Conversely, any partition of the plane into finitely many infinite convex polygonal faces comes from a tree drawing in this way.

Tree drawings with convex faces, such as the one in Figure~\ref{fig:example}, can be visually appealing. For instance, the convexity of the faces makes it easy to visually separate vertices in different subtrees of the tree.  However in terms of the \emph{angular resolution} of a drawing (the minimum angle between any two edges incident on the same vertex~\cite{Mal-STOC-92}) drawings with convex faces may require much sharper angles than nonconvex drawings, and we wish to alleviate this fault by using as wide angles as possible. This motivates the main problem we consider here:

\begin{description}
\item[Tree Drawing With Convex Faces and Optimal Angles:]\hfil\\
Given as input a tree $T$, find a drawing of $T$ with convex faces, having the maximum angular resolution possible among all such drawings.
\end{description}

\begin{figure}[t]
\centering\includegraphics[width=2.25in]{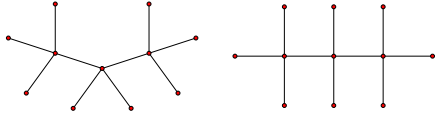}
\caption{Different embeddings of a tree can have different optimal angular resolutions.}
\label{fig:reorder}
\end{figure}

\noindent
We consider two different versions of this problem. In one version a plane embedding of $T$ is given by ordering the edges at each vertex; this ordering must be respected by the drawing. In the other version, the edges at each node of $T$ may be permuted arbitrarily. The optimal angular resolution may differ depending on which version of the problem we consider; for instance, in the tree shown in Figure~\ref{fig:reorder}, the embedding on the left has optimal angular resolution $2\pi/5$ while the right embedding has optimal angular resolution $\pi/2$. Our main results are that the optimal angular resolution, and a drawing achieving that resolution, may be found in linear time, both for the plane case and the unembedded case.

Our tree drawing algorithms set the slopes of all edges in the drawing, before setting the lengths of the edges and placing the vertices. For the slope-setting phase, the definition of having convex faces and the angular resolution do not depend on the choice of root for $T$, so we may assume that initially $T$ is unrooted. However, it will be useful to choose a particular root, depending on the structure of $T$. After the slopes are set, we may return to the original root of $T$ (if it has one) and use that information when we set edge lengths and place vertices.
 
The problem of choosing slopes so that any setting of edge lengths is non-crossing can also be solved by drawings in which the faces are non-convex, as long as paths between consecutive leaves have ranges of angles of at most $\pi$; for instance the nonconvex path on the right of Figure~\ref{fig:arches} is always non-crossing. However, this additional generality does not allow for improved angular resolution, so we restrict our attention to drawings with convex faces henceforth.

\section{Paths and Rakes}

Before describing our main algorithm, we treat some special cases that are problematic for it. These same cases, as subtrees of our input tree, also play a key role in our main algorithm itself.

A \emph{path} is a tree in which all nodes have degree at most two. Clearly, the optimal angular resolution for a drawing of a path with convex faces is $\pi$, achieved by a drawing in which all vertices lie on a common line.

We define a \emph{rake} to be a tree in which all nodes have degree at most three, and in which some path connects all degree-three vertices. Let $T$ be a rake, and let $P$ be a minimal directed path connecting all degree three vertices of $T$.  If $T$ is embedded in the plane, each degree-three vertex~$v$ interior to $P$ has one incoming edge in $P$, one outgoing edge in $P$, and one unoriented edge not belonging to $P$. We say that $P$ makes a \emph{left turn}  at $v$ if the clockwise ordering of these three edges is the incoming edge, the outgoing edge, and the unoriented edge, and we say that $P$ makes a \emph{right turn} at $v$ otherwise. We define a \emph{double turn} to be a pair of consecutive turns in $P$ that are both left or both right.

\begin{figure}[t]
\centering\includegraphics[width=2.75in]{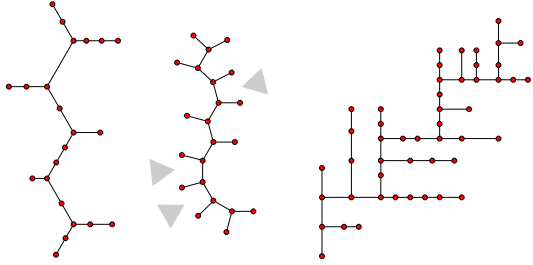}
\caption{Optimal angular resolution drawings of rakes. Left: a rake with no double turns, with angular resolution $2\pi/3$. Center: a rake with three double turns (marked by the gray triangles), requiring angular resolution $7\pi/12$. Right: a construction with angular resolution~$\pi/2$ for any rake.}
\label{fig:rakes}
\end{figure}

\begin{lemma}
\label{lem:rake}
An unembedded rake has optimal angular resolution $\frac{2\pi}3$.
An embedded rake with $k$ double turns has optimal angular resolution
$\pi(\frac12 + \frac1{6+2k})$.
\end{lemma}

\begin{proof}
If a rake has no double turns, it may be drawn with its edges lying on the edges of a tiling of the plane by regular hexagons, as shown in Figure~\ref{fig:rakes}(left), and if the input is an unembedded rake then we may choose an embedding in which the turns alternate left and right and achieve this angular resolution. This is clearly optimal for any tree with degree three nodes.

For an embedded rake with double turns, such as the one in Figure~\ref{fig:rakes}(center), we consider the sequence of angles between consecutive leaves of the tree. These angles must be nonnegative and total $2\pi$. If the angular resolution is $\frac{\pi}{2}+\epsilon$, then the angle between the two paths incident to a degree three node that is not a turn is at least $\frac{\pi}{2}+\epsilon$. The angle between one of these paths and the path incident to the nearest turn is at least $2\epsilon$, because these two paths are connected via two angles of at least $\frac{\pi}{2}+\epsilon$. Similarly, the angle between the two paths in a double turn is at least $2\epsilon$.  Remaining pairs of consecutive leaves may be parallel. Adding all these angles, we get $2(\frac{\pi}{2}+3\epsilon)$ for the angles near the ends of the paths, and $2\epsilon$ for each double turn, for a total of $\pi+(6+2k)\epsilon$. Since this must equal at most $2\pi$, an upper bound on angular resolution of the stated form follows.

To achieve this bound, first assign angles to the leaves of the tree exactly matching the formula above: $\frac{\pi}{2}+\epsilon$ between the two paths incident to a degree three node that is not a turn, etc. The path edges are then assigned angles of $\frac{\pi}{2}+\epsilon$ from the preceding leaf. With this angle assignment, all paths between consecutive leaves are seen to form convex arches, so, as we have already discussed, we may assign edge lengths arbitrarily resulting in a tree drawing with convex faces and the stated angular resolution.
\end{proof}

When $k$ is large, the angular resolution $\pi(1/2 + 1/(6+2k))$ closely approaches $\pi/2$. A very simple construction achieves angular resolution $\pi/2$ for any rake: simply draw the path connecting the degree three nodes by a monotonic path consisting of alternating and horizontal line segments, with the path proceeding horizontally after each right turn and vertically after each left turn, as shown in Figure~\ref{fig:rakes}(right). More generally, the same algorithm shows the following:

\begin{lemma}
\label{lem:subrake}
Let $T$ be a rake, and let two slopes $\theta_1$ and $\theta_2$ be given.
Then $T$ may be drawn with all edges having slopes $\theta_1$ or $\theta_2$,
so that all faces of $T$ except the outer face at its root are convex.
\end{lemma}

When $|\theta_1-\theta_2|\le\pi/2$, the angular resolution of the drawing produced by Lemma~\ref{lem:subrake} is $|\theta_1-\theta_2|$.
We will use this construction as part of our algorithm for drawing trees that are not rakes.

\section{Triple Rakes}

\begin{figure}[t]
\centering\includegraphics[width=1.75in]{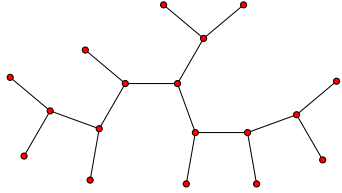}
\caption{A triple rake with one double turn and one short path. The optimal angular resolution for this tree as embedded is~$5\pi/18$.}
\label{fig:triplerake}
\end{figure}

Given a tree $T'$, in which the maximum vertex degree is three, let $T'$ be the minimal spanning subtree of the degree three vertices in $T$. $T$ is a rake if and only if $T'$ is a path, but the next simplest case is when $T'$ contains a single degree three vertex $t$.  In this case, if we root $T$ at $t$, the three subtrees descending from $t$ are rakes, so we call $T$ a \emph{triple rake} (Figure~\ref{fig:triplerake}).

If $T$ is embedded, then in each of the three paths of $T'$, oriented from $t$ to a leaf, we may define left turns, right turns, and double turns as we did in rakes. Additionally, we define a \emph{short path} in $T'$ to be a path with no turns; that is, a single degree three vertex connected by paths to $t$ and to two leaves.

\begin{lemma}
\label{lem:triplerake}
If $T$ is an unembedded triple rake with $s$ short paths, its optimal angular resolution is
$\pi(\frac12+\frac1{2(9-2s)})$. If $T$ is an embedded triple rake with $s$ short paths and $d$ double turns, its optimal angular resolution is $\pi(\frac12+\frac1{2(9-2s+2d)})$.
\end{lemma}

\begin{proof}
The unembedded case follows from the embedded case, by embedding the tree with no double turns.
For the embedded case, we assume that the optimal angular resolution is $\pi/2+\epsilon$, and count the number of times the angles increase by $\epsilon$ as we progress around the leaves of a drawing of the tree, as we did in Lemma~\ref{lem:rake}.  The angle between the two bottom leaves of a rake is at least $\pi/2+\epsilon$, and in a rake that does not come from a short path the angle between one of the two bottom leaves and the next nearest path is another $2\epsilon$.  Additionally, as in Lemma~\ref{lem:rake}, each double turn leads to an angle increase of $2\epsilon$. The total angle increase as we go around the tree is $3\pi/2 + (9-2s+2d)\epsilon=2\pi$, from which the bound follows. This analysis fixes the angles of all leaves of the tree, from which it is straightforward to find a drawing achieving the stated bound.
\end{proof}

\section{Plane Trees}

We are ready to describe our general bound for the optimal angular resolution of a plane embedded tree. We assume our tree $T$ is not a path, rake, or triple rake, as those special cases were handled in previous sections. As the problem of tree drawing with convex faces does not depend on the root of the tree, we choose a new root $r$ as follows:

\begin{itemize}
\item If $T$ contains a vertex incident to four or more edges, we let $r$ be any such vertex.
\item Otherwise, let $T'$ be the minimal subtree of $T$ containing all degree-three vertices in $T$; $T'$ can be formed by removing from $T$ all leaves of $T$, and all paths of degree-two vertices in $T$ that lead to a leaf. $T'$ cannot be a path, as we have assumed that $T$ is not a rake. Therefore, there exists a vertex in $T'$ with degree three. We let $r$ be any such vertex.
\end{itemize}

Once we have rooted $T$ at $r$, we consider for each node $v$ and each child $w$ of $v$
the subtree $T_w$ formed by $v$, $w$, and all descendants of~$w$. It will be important for our algorithms to be able to determine whether $T_w$ is a path or rake.

\begin{lemma}
For all $w$ we can determine whether $T_w$ is a path, a rake, or a tree that is not a path or rake, in total time $O(n)$.
\end{lemma}

The algorithm for performing this determination is a simple bottom-up calculation on $T$; we omit the details.

\begin{figure}[t]
\centering\includegraphics[width=3.5in]{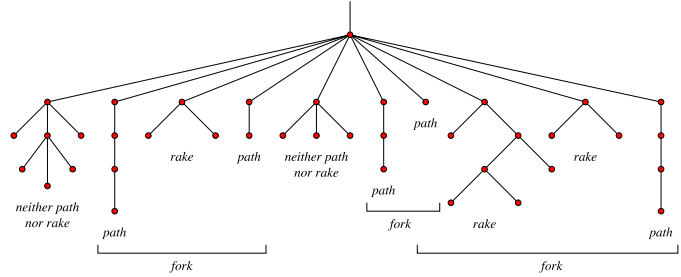}
\caption{A vertex with three forks. The subtrees descending from the top vertex of the figure contain additional forks.}
\label{fig:forks}
\end{figure}

We define a \emph{fork at $v$} to be a subsequence of two or more children $w_i$ of~$v$, contiguous in the ordering of the children given by the plane embedding of~$T$ such that the trees $T_{w_i}$ for the first and last child in the subsequence are paths and all intermediate trees are rakes (Figure~\ref{fig:forks}). We can also identify a fork with a subtree, formed by $v$, the subsequence of children in the fork, and all descendants of those children. When $v$ is not the root, the sequence of children of $v$ is a linear order, but for the root $v=r$ we consider this sequence as a cyclic order and allow any linear subsequence of this order. In particular, when $r$ has one child forming a path subtree and all its other children form rakes, we consider the sequence starting at the path, continuing through all the rakes, and ending at the path again to form a fork.

\begin{lemma}
\label{lem:two-forks}
If $T_w$ is a rake, it contains exactly one fork, at the bottommost vertex with two children. Otherwise $T_w$ contains at least two forks at a single vertex.
\end{lemma}

\begin{proof}
We use induction on the height of $T_w$. If $w$ has one child $x$, and $T_w$ is not a rake, then $T_x$ is also not a rake, and the result follows. If $w$ has two children, and $T_w$ is not a rake, then either one child $x$ is not a rake, and again the result follows, or both children are rakes, and we have one fork in each.
If $w$ has three or more children, one of which is neither a path nor a rake, then we have two forks in that child alone. If there are two or more rakes, then again we have one fork in each. If there is only one rake descending from $w$, then the other two subtrees descending from $w$ are paths, and we have one fork at $w$ itself and one in the rake. Finally, if all descendants of $w$ form paths, then there are at least two forks at~$w$.
\end{proof}

\begin{lemma}
\label{lem:fork-angle}
Let $F$ be a fork at a vertex $v$, containing $r$ rakes, in a tree drawing with convex faces and angular resolution $\theta$. Then the angle between the first leaf and the last leaf in $F$ is at least $(r+1)\theta$.
\end{lemma}

\begin{proof}
That angle is needed just for the $r+2$ edges connecting $v$ to its children in the fork. Adding the remaining edges in the fork cannot decrease the angle any further.
\end{proof}

\begin{lemma}
\label{lem:plane-ub}
Let $T$ be a tree containing $f$ forks, as rooted at $r$. Then any drawing of $T$ with convex faces has angular resolution at most $2\pi/f$.
\end{lemma}

\begin{proof}
We prove more generally that, if $T$ has $f$ forks at some node $v$ or its descendants, in a drawing with angular resolution $\theta$, then the angle between the first and last leaves descending from $v$ is at least $f\theta$. The result comes from applying this bound to the trees descending from the children of the root.

To prove a lower bound of $f\theta$ on the angle between the first and last leaves descending from $v$, we use induction on the height of the subtree. The sequence of slopes of leaves increases monotonically as we proceed clockwise around the tree, increasing (by induction) by $\theta$ times the number of forks in each subtree that is not a path or rake.  In a rake that is not part of a fork, the bottommost two paths must have an angle of at least $\theta$, matching the single fork (Lemma~\ref{lem:two-forks}) that exists in the rake. Finally, each fork at $v$, containing $r$ rakes, leads to a total of $(r+1)$ forks when we include the fork within each rake, and by Lemma~\ref{lem:fork-angle} the increase in angle in this case again matches the number of forks.
\end{proof}

\begin{lemma}
\label{lem:four-corners}
In a tree that is not a path, rake, or triple rake, rooted at $r$ as described above, there are at least four forks in the tree.
\end{lemma}

\begin{proof}
If two or more of the trees descending from children of $r$ are not paths or rakes, the result follows from Lemma~\ref{lem:two-forks}. If $r$ has four or more children, exactly one of which is not a path or rake, then the other three children have two rakes, or form a fork with one rake, or form two forks; in all cases there are two forks from the non-rake child and two from the other three children. If $r$ has four or more children, all of which are paths or rakes, then each path has a fork clockwise of it and each rake has a form inside it, so again the total number of forks is at least four. Finally, if $r$ has degree three, then (by our choice of root) none of the subtrees descending from it is a path, and (since the tree is not a triple rake) at least one of these subtrees is not a rake, so we get two forks from this non-rake subtree and one each from the other two subtrees.
\end{proof}

By Lemmas~\ref{lem:plane-ub} and~\ref{lem:four-corners}, the angular resolution of a tree that is not a path, rake, or triple rake is at most $\pi/2$, so we may use the construction of Lemma~\ref{lem:subrake}.

\begin{lemma}
\label{lem:plane-lb}
Let $T$ be a tree containing $f$ forks, as rooted at $r$. Then $T$ has a drawing with convex faces and angular resolution $2\pi/f$.
\end{lemma}

\begin{proof}
We assign slopes to the edges of $T$ in postorder. In a subtree containing $f'$ forks, the angle from the first leaf to the last leaf will be $2\pi f'/f$; thus, the total angle around the entire tree will be $2\pi$ as desired. When assigning slopes to the edges of the subtree rooted at $v$, we consider the children of $v$ in order. 

Each subtree that is not a path or rake has its first leaf slope equal to that of the leaf immediately preceding the tree, and by induction can be drawn with the stated angle bound. We choose the slope of the edge leading from $v$ to the subtree in such a way that it bisects the angle formed by the subtree's first and last leaves; in this way, the angle between two consecutive edges incident to $v$ is half the angle spanned by the two subtrees, and thus at most $\pi$.  In addition, this choice of slope is guaranteed to be at least $2\pi/f$ greater than that of the first leaf in the subtree, and at least $2\pi/f$ less than that of the last leaf in the subtree, so each edge incident to $v$ will form an angle of at least $2\pi/f$ with the preceding and succeeding edges. Finally, we can show that (together with our choice of slopes for other types of subtree) this choice of root slope will always be within the range of slopes of the edges at the child vertex of the subtree.

Each path that is not the second path of a fork is given a slope equal to that of the previous leaf. Each rake that is part of a contiguous sequence of rakes following a path is drawn using Lemma~\ref{lem:subrake} in a way that increases the slope by $2\pi/f$, matching the bound for a subtree with a single fork; we align the root edge of the rake with the slope of its last leaf. Each other rake is drawn by Lemma~\ref{lem:subrake} with its root edge aligned with the slope of its first leaf. Finally, the second path of any fork is assigned a slope greater than the preceding leaf by $2\pi/f$.
\end{proof}

Putting these lemmas together, we have the following result:

\begin{theorem}
\label{thm:plane}
Let $T$ be an unrooted plane tree. Then in time $O(n)$ we can find a drawing of $T$ with convex faces and optimal angular resolution.
\end{theorem}

\begin{proof}
We first test whether $T$ is a path, rake, or triple rake.
If it is a path, we embed it trivially with angular resolution $\pi$. If it is a rake, we apply Lemma~\ref{lem:rake}, and if it is a triple rake, we apply Lemma~\ref{lem:triplerake}. Otherwise, we root $T$ as described at the start of this section, determine which subtrees are paths and rakes, count forks at each node, and apply the drawing method described in Lemma~\ref{lem:plane-lb}. The optimality of the angular resolution of this method follows from Lemma~\ref{lem:plane-ub}.
\end{proof}

\section{Unembedded Trees}

We are now ready to handle trees that do not have a plane embedding already fixed. Our choice of root at the beginning of the previous section does not depend on the embedding, so we may use it without change. However, the definition of a fork depends strongly on the embedding. Our goal in finding an embedding maximizing the angular resolution is to minimize the number of forks. To do so, define the \emph{excess} of a node $v$ that is not the root of the tree to be $E_v=\max(0,P_v-N_v-1)$, where $P_v$ is the number of trees descending from $v$ that are paths and $N_v$ is the number of trees descending from $v$ that are neither paths nor rakes. For the root $r$, define $P_r$ and $N_r$ similarly, and let $E_r=\max(0,P_v-N_v)$.  The \emph{total excess} $E(T)$ is the sum of the excesses at each node.

\begin{lemma}
\label{lem:minforks}
Every plane embedding of tree $T$ has at least $E(T)$ forks.
There exists an embedding of $T$ with exactly $E(T)$ forks.
\end{lemma}

\begin{proof}
In any embedding, at node $v$ there are $P_v$ paths, forming $P_v-1$ potentially-consecutive pairs of paths ($P_v$ pairs in the cyclic order at the root).
At most $N_v$ of these pairs can be separated by a subtree that is not a path or a rake, and the remaining pairs form forks, so there are at least $E_v$ forks at $v$ and $E(T)$ overall.
This bound may be achieved by, at each node, placing the paths and the trees that are not paths or rakes in alternating order for as many alternations as possible. The placement of rakes in the ordering at each vertex and the ordering of the descendants within each rake does not affect the number of forks in~$T$.
\end{proof}

\begin{theorem}
Let $T$ be an unrooted unembedded tree.  An embedding of $T$ and its drawing with convex faces and optimal angular resolution can be found in $O(n)$ time.  If $T$ is not a path, rake, or triple rake, the optimal angular resolution among all drawings of $T$ with convex faces is $2\pi/E(T)$.
\end{theorem}

\begin{proof}
We first test whether $T$ is a path, rake, or triple rake.
If it is a path, we embed it trivially with angular resolution $\pi$. If it is a rake, we apply Lemma~\ref{lem:rake}, and if it is a triple rake, we apply Lemma~\ref{lem:triplerake}. Otherwise, we root $T$ as described at the start of the previous section, embed it with the minimum number of forks via Lemma~\ref{lem:minforks}, and apply Theorem~\ref{thm:plane} to the resulting embedded tree using the same root. The bound on the angular resolution follows from Lemmas~\ref{lem:plane-ub}, \ref{lem:plane-lb} and~\ref{lem:minforks}.
\end{proof}

\section{Setting the Lengths}

\begin{figure}[t]
\centering\includegraphics[height=1.85in]{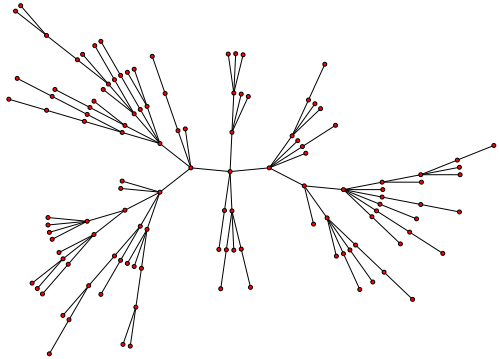}
\caption{The same tree as Figure~\ref{fig:example}, drawn with all edge lengths equal.}
\label{fig:equal}
\end{figure}

Although our focus has been on the edge slopes for tree drawings, we can not produce an actual drawing without setting the edge lengths.  Our slopes have been chosen specifically to allow any possible setting of edge length, without introducing a crossing, so we may choose these lengths arbitrarily, to advance some aesthetic criterion for the drawing or convey some additional information about the tree. We discuss three possible choices of edge length.

\begin{figure}[t]
\centering\includegraphics[height=1.3in]{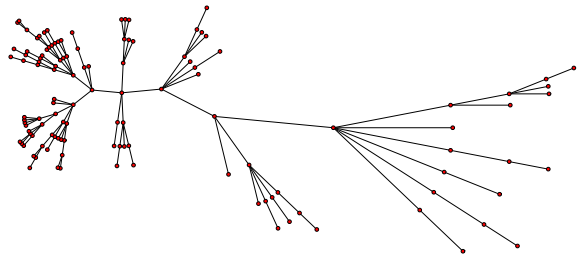}
\caption{The same tree as Figure~\ref{fig:example}, drawn with edge length inversely proportional to the distance from the root.}
\label{fig:invdist}
\end{figure}

\begin{figure}[t]
\centering\includegraphics[height=1.3in]{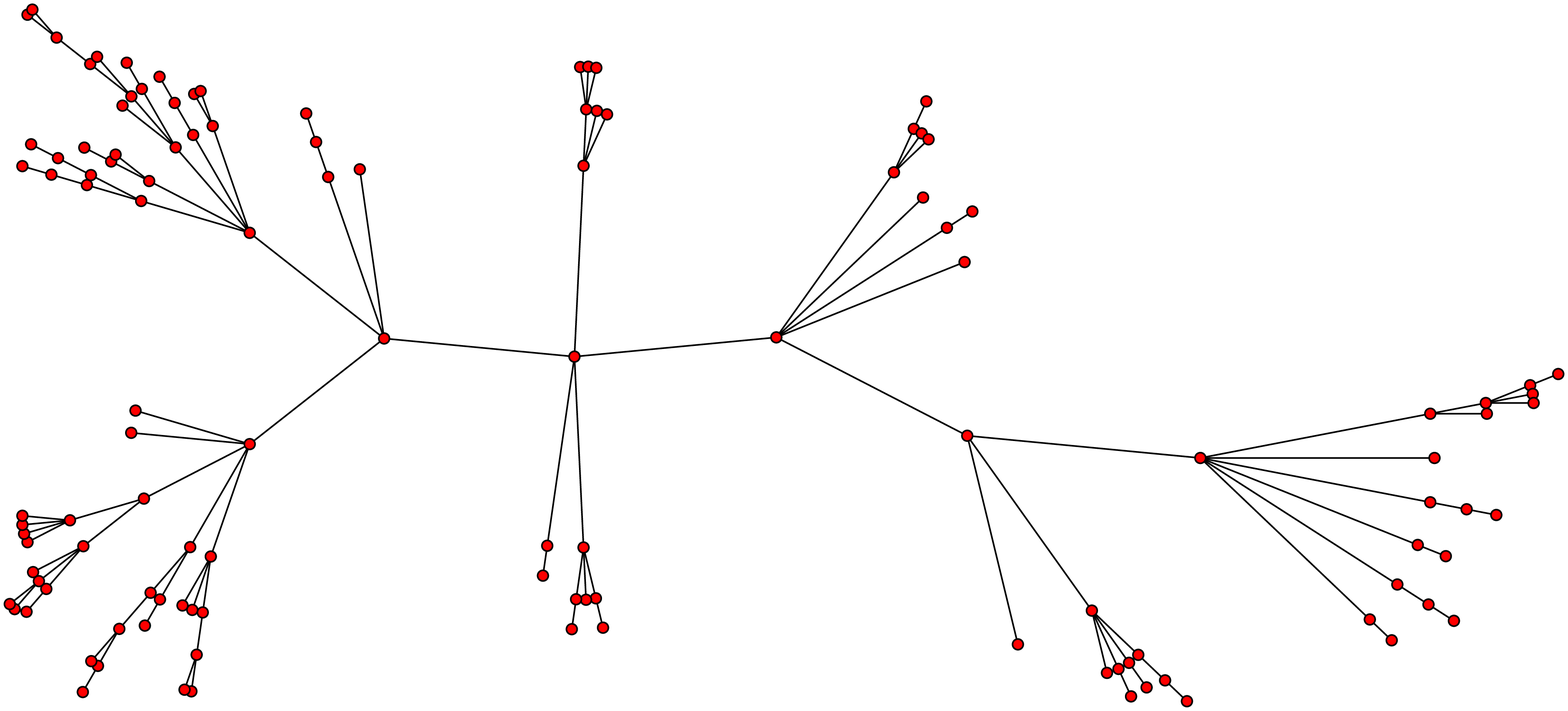}
\caption{The same tree as Figure~\ref{fig:example}, drawn with edge length proportional to the square root of the number of descendants of the edge's top vertex.}
\label{fig:sqrtdesc}
\end{figure}

\begin{description}
\item[Uniform Edge Lengths:]\hfil\\
Assigning all edges the same length can produce a pleasantly uniform vertex spacing. Since all vertices are treated equivalently, this type of drawing is appropriate for unrooted trees.
Trees with uniform edge lengths are shown in Figures~\ref{fig:reorder}, \ref{fig:rakes}(center), \ref{fig:triplerake}, and~\ref{fig:equal}.

\medskip
\item[Radial Drawing:]\hfil\\
We may place the root of a rooted tree at the center of a system of concentric circles, and place each vertex on the circle corresponding to its distance from the root. To do so, we set the lengths of the edges in preorder. When setting the length of an edge $(u,v)$, the placement of $u$ on its circle will already have been fixed. Since the circle for $u$ is inside the circle for $v$, a ray from $u$ in the direction of the slope for edge $(u,v)$ will intersect the circle for $v$ in a unique point, at which we place $v$. We then set the edge length to the distance between the placements of its endpoints.
A drawing in this style is shown in Figure~\ref{fig:example}.
This style of tree can be effective in making visually apparent the root of the tree and the distance of each node from the root, especially if the drawing is displayed with the concentric circles visible, as they are in the figure. The root used for this placement algorithm need not be the same as the root $r$ that was chosen in our angle-setting algorithm.

\medskip
\item[Position in Tree:]\hfil\\
Similarly to radial drawing, we may choose edge lengths that are functions of the position of the edge in the tree. Two drawings of this type are shown in Figures~\ref{fig:invdist} and~\ref{fig:sqrtdesc}, with lengths that are functions of the distance from the root and the size of the subtree rooted at the top vertex of the edge respectively. It may also be of interest to vary edge lengths of this type continuously, to morph between multiple viewpoints in the same tree.

\medskip
\item[Strength of Connection:]\hfil\\
Since the lengths of edges may be arbitrary, we may use them to convey additional information about the tree being drawn. For instance, if some edges form stronger connections between their vertices than others, we may display this by placing more strongly connected vertices closer together, and more weakly connected vertices farther apart.
\end{description}

\section{Dominance Drawing and Downward Drawing}

We have considered drawings that allow the edges to be drawn with any slope. However, it may be of interest to require that no edge from a parent to child is directed upwards (\emph{downward drawing}) or that no edge is directed either upwards or to the right (necessary but not sufficient for \emph{dominance drawing}). Similar techniques to the ones here, using a  linear rather than circular ordering at the root of the tree, can find drawings of these types, with convex faces (except, in the case of dominance drawing, for the face above and to the right of the root) and optimal angular resolution $\theta/f$ where $\theta$ is the range of allowed edge slopes and $f$ is the number of forks in the tree.  We omit the details due to lack of space.

\section{Future Work}

We have shown how to draw trees with convex faces and optimal angular resolution.  It is natural to extend such drawing methods to other graphs. It seems likely that similar methods can optimize the angles of  \emph{pseudotrees} (connected undirected graphs with a single cycle) drawn so the cycle forms a bounded convex face and all other faces are convex and unbounded. Similarly, a reviewer suggested {\em Halin graphs}, are formed from a tree with no degree-two vertices by connecting its leaves into a cycle; we can form convex drawings of such graphs by choosing edge lengths for the tree so that the leaves are in convex position, but it is not clear how to optimize the angular resolution of such drawings.

Another reviewer suggested that it may be preferable to draw paths of degree-two vertices with small bends between the edges rather than having all edges lie on the same straight line, in order to make the vertex positions more apparent. Such a preference could be quantified by a modified definition of angular resolution, for instance one that measured the angle between any two consecutive edges as the smallest of the two complementary angles formed by the edges' lines; we expect that techniques similar to ours could be used to optimize this modified angular resolution, although the details would differ and the faces of the resulting drawings would likely be nonconvex.

We may also seek stronger optimality conditions for our drawings. In particular, consider the vector of angles between consecutive edges in a tree drawing, sorted from smallest to largest. Our algorithm finds a vector with first coordinate as large as possible, but can we find the maximum possible vector in the lexicographic ordering of all possible vectors?  For plane trees, it seems possible to solve this problem in polynomial time by setting up a sequence of linear programs to optimize each successive coordinate of the vector of angles, but this is neither efficient nor satisfactorily combinatorial. For unembedded trees, a solution to the lexicographic optimization problem seems even more difficult, and we do not know whether it is likely to be polynomial or NP-hard.

\raggedright
\bibliographystyle{abbrv}
\bibliography{tree-angles}

\end{document}